\documentclass[]{aastex}

\usepackage{emulateapj5}
\usepackage{onecolfloat}
\usepackage{graphicx} 
\usepackage{fancyheadings} 
\usepackage{ulem}
\usepackage{rotating}
\usepackage{lscape}

\newcommand{\kms}{km~s$^{-1}$ }
\newcommand{\cm}[1]{\, {\rm cm^{#1}}}
\newcommand{\mkms}{{\rm \; km\;s^{-1}}}
\newcommand{\delv}{\Delta v}

\newcommand{\lya}{Ly$\alpha$}

\newcommand{\N}[1]{{N({\rm #1})}}
\newcommand{\sci}[1]{{\rm \; \times \; 10^{#1}}}
\newcommand{\qlla}{SDSS0927+5621}
\newcommand{\qllb}{SDSS0953+5230}
\newcommand{\slla}{SLLS0927+5621}
\newcommand{\sllb}{SLLS0953+5230}
\newcommand{\mnhi}{N_{\rm HI}}
\newcommand{\mnciv}{N_{\rm CIV}}
\newcommand{\nhi}{$N_{\rm HI}$}

\def\mfnhi{f_{\rm{HI}} (\mnhi)}

\def\nhi{$N_{\rm HI}$}

\def\omt{$\Omega_m^{Total}$}
\def\momt{\Omega_m^{Total}}

\begin{document}

\twocolumn[%
\submitted{Submitted to ApJL; Revised June 22 2006}

\title{Super-Solar Super Lyman Limit Systems}

\author{
Jason X. Prochaska\altaffilmark{1,2},
John M. O'Meara\altaffilmark{3}, 
St\'ephane Herbert-Fort\altaffilmark{2,4}
Scott Burles\altaffilmark{3}, 
Gabriel E. Prochter\altaffilmark{1,2}, 
and Rebecca A. Bernstein\altaffilmark{5}
\\}

\begin{abstract}
We present abundance measurements for two super Lyman Limit systems
(SLLS; quasar absorption line systems with 
$10^{19} \cm{-2} < \mnhi < 10^{20.3} \cm{-2}$) selected from a 
set of metal-strong absorbers in the Sloan Digital Sky Survey quasar
database.  After applying estimate corrections 
for photoionization effects, we derive
gas-phase metallicities of $\lbrack$M/H$\rbrack$=$+0.7 \pm 0.2$\,dex
for the SLLS at $z=1.7749$ toward \qlla\ and 
$\lbrack$M/H$\rbrack$=$+0.05 \pm 0.1$\,dex for the SLLS at $z=1.7678$ toward \qllb.
The former exhibits among the highest gas metallicity 
of any astrophysical environment and its total metal surface density
exceeds that of nearly every known damped \lya\ system.
The properties of these absorbers -- high metallicity
and large velocity width ($\delv > 300 \mkms$) --
resemble those of gas observed in absorption in
the spectra of bright, star-forming galaxies at high redshift. 
We discuss the metal mass density of the SLLS based on these
observations and our ongoing SLLS survey and argue that a
conservative estimate to the total metal budget at $z=2$ is greater than
$15\%$ of the total, suggesting that
the metal-rich LLS may represent the dominant metal
reservoir in the young universe.

\end{abstract}
]

\altaffiltext{1}{Department of Astronomy and Astrophysics, UCO/Lick Observatory, 
University of California, 1156 High Street, Santa Cruz, CA 95064}
\altaffiltext{2}{Visiting Astronomer, W.M. Keck Observatory which is a joint 
facility of the University of California, the California Institute of Technology, 
and NASA}
\altaffiltext{3}{MIT Kavli Institute for Astrophysics and Space Research,
Massachusetts Institute of Technology, 77 Massachusetts Avenue, Cambridge MA 02139}
\altaffiltext{4}{University of Arizona/Steward Observatory, 933 N Cherry Avenue,
Tucson, AZ 85721}
\altaffiltext{5}{Department of Astronomy, University of Michigan, 
	Ann Arbor, MI 48109}

\pagestyle{fancyplain}
\lhead[\fancyplain{}{\thepage}]{\fancyplain{}{PROCHASKA ET AL.}}
\rhead[\fancyplain{}{Super-solar Super LLS}]{\fancyplain{}{\thepage}}
\setlength{\headrulewidth=0pt}
\cfoot{}

\section{Introduction}

With the recent successes of high redshift galaxy surveys, the star
formation history of the young universe is revealing itself
\citep{madau96,goods04}.  Although debate 
continues on various aspects of the measurements
(e.g.\ dust extinction, sample selection, sample variance),
it is evident that the star formation rate at $z>2$ was substantially
higher than the current epoch.  
Accordingly, the massive stars which light up these galaxies must
have produced copious metals and redistributed them via supernovae
to the galaxy and surrounding medium.  A valuable test of this
picture, therefore, is to obtain an accurate census of metals at
$z \approx 2$ and compare against the amount predicted by the
integrated the star formation history \citep{pagel02,pettini04,bouche05I}.
In the following, we will adopt a metal mass density at $z=2$
of $\Omega_m^{Total} = 3 \sci{-5} h_{70}^{-2}$ based on the work
of \cite{bouche06II}.

Prior to these galaxy surveys, 
analysis of heavy metals in quasar absorption line (QAL) systems --
the gas within and in between high $z$ galaxies --
demonstrated that star formation was ubiquitous
at $z>2$ \citep{tytler95,wolfe94,pettini94}.  
Metals are present in excess of the
primordial value throughout much of the universe 
\citep{schaye03,simcoe04,bergeron05}
and a direct census of the intergalactic medium (IGM) 
adds up to $5-15\%$ of \omt \citep{pettini04,schaye03}.
Regarding the interstellar medium of galaxies, 
the damped \lya\ systems (QAL systems
with $\mnhi \geq 2\sci{20} \cm{-2}$) offer the most direct
means of measurement \citep{pettini94,pro_mtl03}.  
Surveys of these absorption systems have demonstrated that the
metals associated with \ion{H}{1} gas in DLAs
contribute as comparable a fraction as the IGM probed through
the \lya\ forest.
The metals in stars can be estimated from the luminosity functions
and metallicity estimates of galaxy surveys; 
\cite{bouche06II} estimates this contributiotn to be $20-50\%$ at $z=2$.
Altogether the census of metals in the IGM, the ISM probed by DLA, and stars
falls short by as much as 70\% of \omt.

At present, there are several suggested explanations for this discrepancy.
These include: (1) a significant
portion of the metals
are within dusty neutral gas (possibly molecular) which obscures
any background quasar avoiding detection \citep{fall93,vladilo05},
(2) a significant fraction of metals are sequestered within a hot,
diffuse, collisionally ionized gas,
which is difficult to probe with rest-frame ultraviolet transitions
\citep{ferrara05}, (3) metals (primarily O~VI) are locked in warm--hot gas 
\citep{simcoe02}, or (4) metals are located in high metallicity ``feedback'' 
systems \citep{simcoe06}.
All of these phases are certain to contribute, some to a larger extent 
than others, and it is possible
that a complete solution will include significant contributions
from each.

In this paper, we wish to highlight another reservoir of metals:
highly photoionized, yet optically thick absorption systems.  
Specifically, we wish to examine the 
possible contribution from the Lyman Limit systems (LLS; absorbers
with $\mnhi > 10^{17.2} \cm{-2}$ and restricted here to have 
$\mnhi < 10^{20.3} \cm{-2}$),
the heretofore neglected sibling of the QAL systems.
While the intergalactic medium examines gas arising in regions
with overdensity $\delta \equiv \rho/\bar\rho < 10$ and the
damped \lya\ systems describe gas in overdense ($\delta > 200$)
galactic regions,  the LLS are likely to 
represent the interface between dense galactic and tenuous intergalactic
gas.  It is reasonable to speculate that these absorbers
probe gas related to galactic feedback processes and could therefore
be a significant metal reservoir.
The IGM and DLA systems have been surveyed extensively for 
metals \citep{schaye03,simcoe04,pro_mtl03} but analysis of the
LLS has been limited to a few systems \citep{prochaska99,pb99}
and a small survey of super Lyman limit systems (SLLS), systems with 
$\mnhi = 10^{19} - 10^{20.3} \cm{-2}$ \citep{mirka03,peroux_slls03,peroux05}.
Because even the SLLS outnumber DLA by approximately a factor of four,
the LLS could easily contribute a significantly larger
metal mass than the DLA.  

To make a direct comparison, one must
consider the frequency distribution, ionization state, and mean
metallicity of the LLS, all of which are poorly constrained.
In this Letter, we will consider the current constraints and 
highlight the prospective importance of the SLLS. 
\cite{peroux06} have recently presented measurements on a low
redshift SLLS which shows a super-solar metallicity\footnote{One notes
that \cite{rtn06} measured this absorber to have $\log \mnhi = 20.5$
and therefore identified it as a DLA system.  Our analysis of the 
HST/STIS spectra gives $\log \mnhi = 20.3 \pm 0.2$.} and the authors
proposed that similar absorbers at high redshift could contain 
an important fraction of the metal budget.  Here, we report on 
two super-solar SLLS at $z_{abs}=1.8$ drawn from our survey of 
metal-strong DLA candidates \citep{herbert06} and an on-going survey
of LLS.  
The more extreme of the two systems exhibits greater than 
5 times solar metallicity and has a metal surface density which 
matches all of the DLA with 
metallicity measurements at $z=1.6$ to 2.2.
We will argue that the LLS are likely to account for at least $15\%$
of the metal budget at high redshift and possibly the remainder of
`missing' metals.

\begin{figure}[ht]
\begin{center}
\includegraphics[width=3.5in]{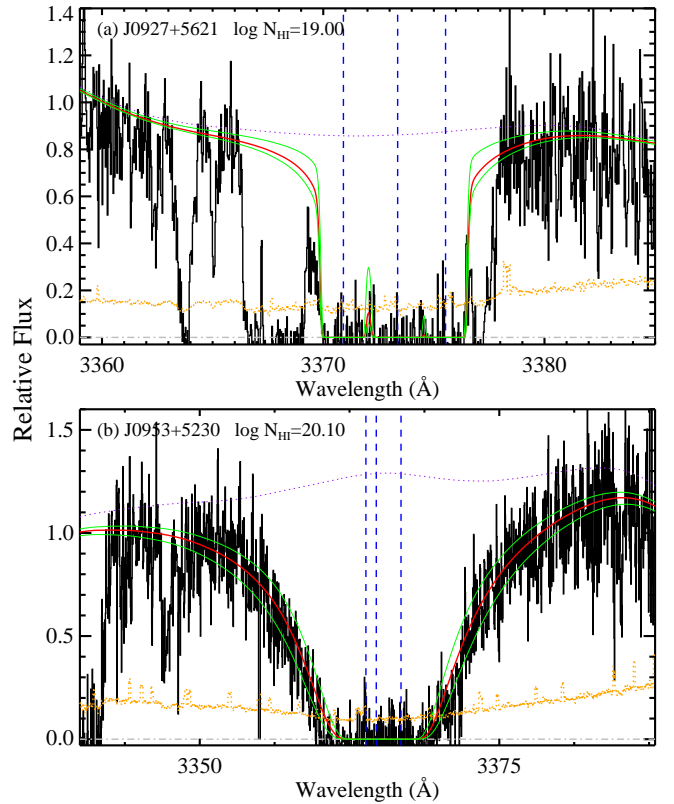}
\end{center}
\caption{\lya\ profiles of the SLLS at (a) $z=1.775$ toward \qlla\
and (b) $z=1.768$ toward \qllb.  The dotted curves indicate our
estimate of the quasar continuum convolved with the blaze function
of the HIRES spectrometer.  The dashed vertical lines indicate the
redshift of `clouds' included in our best-fit solution (solid red line).
}
\label{fig:lya}
\end{figure}

\section{Observations and Chemical Abundances}

\cite{herbert06} have recently published a survey for QAL systems
with very large metal-line column densities (termed metal-strong DLA
candidates) from the 
Sloan Digital Sky Survey database \citep{sdssdr3}.
The overwhelming majority of these absorbers are metal-strong
damped \lya\ systems (MSDLA), yet follow-up observations have revealed
that a small fraction have $\mnhi < 2\sci{20} \cm{-2}$.
Two examples include the SLLS at $z \approx 1.8$ toward quasars
SDSSJ0927+5621 and \\
SDSSJ0953+5230 (hereafter, \slla\ and \\
\sllb).

We observed the quasars SDSSJ0927+5621 and \\
SDSSJ0953+5230 on UT 13 April 2005
and 17 March, 2005 for total exposure times of 7200s and 8400s
respectively with the upgraded HIRES spectrometer
\citep{vogt94} on the Keck\,I telescope.  In each case, we
employed the C5 decker ($1.1''$ wide) giving a 
FWHM$\approx 8 \mkms$ resolution and chose a cross-disperser angle which
gave wavelength coverage $\lambda \approx 3150 - 6000$\AA.
We reduced the 2D images with the HIRES Redux 
pipeline\footnote{http://www.ucolick.org/$\sim$xavier/HIRedux/index.html}
and extracted, coadded and normalized the 1D spectrum with the software.
The spectra have a signal-to-noise ratio of $\approx 7$ per 2.6\kms\ pixel
at 3400\AA.  

\begin{figure*}
\begin{center}
\includegraphics[width=6.8in]{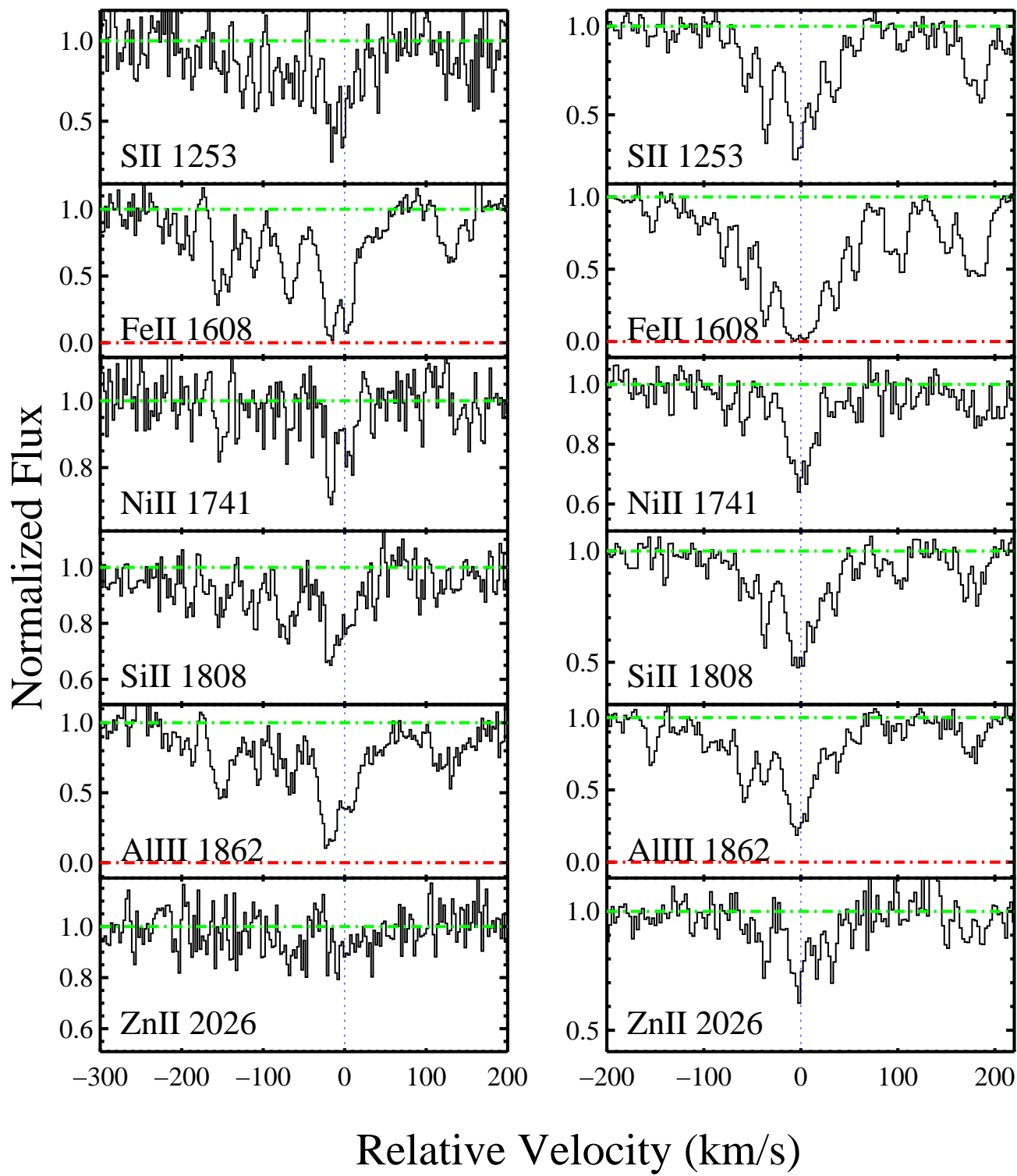}
\end{center}
\caption{Velocity profiles of a subset of the metal-line transitions
observed for the SLLS at LHS: $z=1.7749$ toward \qlla\ and RHS: $z=1.7678$
toward \qllb.}
\label{fig:velp}
\end{figure*}

We have measured the \ion{H}{1} column density of these SLLS by
fitting Voigt profiles to the \lya\ transitions (Figure~\ref{fig:lya}).
For \sllb\ (Figure~\ref{fig:lya}b),
damping wings are obvious and a precise evaluation of
\nhi\ is straightforward, albeit subject to uncertainty dominated
by continuum placement.  In the case of \slla, the absence of
significant damping wings limits $\mnhi < 10^{19.2} \cm{-2}$.
In Figure~\ref{fig:lya}a, we present a fit constructed by distributing
the neutral hydrogen according to the observed metal-line profiles.
The best-fit total hydrogen column density is $10^{19.0} \cm{-2}$,
but we caution that a significantly lower column density is permitted
by the data.
Figure~\ref{fig:velp} presents a subset of the metal-line
transitions observed for the two absorbers.  We have measured
column densities for all the transitions using the apparent 
optical depth method \citep{savage91} and list the values and
errors in Table~1.  At HIRES resolution, the profiles
are well resolved and there is no indication of `hidden' line saturation
from column density measurements of multiple transitions of the same ion.

\begin{table*}[ht]\footnotesize
\begin{center}
\caption{{\sc IONIC COLUMN DENSITIES\label{tab:colm}}}
\begin{tabular}{lccccccc}
\tableline
\tableline
Ion & $\lambda$ & & J0927+5621 & & & J0953+5230 \\
&& flg$^a$
& $\log N$ & $\log N_{adopt}^b$
& flg$^a$ & $\log N$ & $\log N_{adopt}^b$ \\
\tableline
\ion{H}{1} & 1215.67 & 1 & & 19.00$^{+0.10}_{-0.25}$& 1 & & 20.10$^{+0.10}_{-0.10}$ \\
\ion{C}{2} & 1334.532 & 3&$> 15.49$&$> 15.49$&3&$> 15.62$&$> 15.62$\\
\ion{C}{4} & 1548.195 & 2&$> 15.31$&$> 15.52$&2&$> 15.22$&$> 15.31$\\
\ion{C}{4} & 1550.770 & 2&$> 15.52$&&2&$> 15.31$&\\
\ion{N}{1} & 1199.550 &  & & &3&$> 14.92$&$> 14.92$\\
\ion{N}{5} & 1242.804 &  & & &0&$ 13.33 \pm 0.11$&$ 13.33 \pm 0.11$\\
\ion{O}{1} & 1302.168 & 3&$> 15.71$&$> 15.71$&3&$> 15.86$&$> 15.86$\\
\ion{O}{1} & 1355.598 &  & & &5&$< 17.91$&\\
\ion{Al}{2} & 1670.787 & 3&$> 14.02$&$> 14.02$&3&$> 14.13$&$> 14.13$\\
\ion{Al}{3} & 1854.716 & 2&$> 13.93$&$ 14.05 \pm 0.01$& & & $ 13.86 \pm 0.01$\\
\ion{Al}{3} & 1862.790 & 0&$ 14.05 \pm 0.01$&&0&$ 13.86 \pm 0.01$&\\
\ion{Si}{2} & 1264.738 & 4&$< 12.61$&$ 15.58 \pm 0.02$&4&$< 11.97$&$ 15.67 \pm 0.01$\\
\ion{Si}{2} & 1304.370 & 3&$> 15.37$&&3&$> 15.46$&\\
\ion{Si}{2} & 1526.707 & 3&$> 15.23$&&3&$> 15.34$&\\
\ion{Si}{2} & 1808.013 & 1&$ 15.58 \pm 0.02$&&1&$ 15.67 \pm 0.01$&\\
\ion{Si}{3} & 1206.500 & 2&$> 14.41$&$> 14.41$&2&$> 14.46$&$> 14.46$\\
\ion{Si}{3} & 1892.030 & 4&$< 16.70$&&4&$< 16.39$&\\
\ion{Si}{4} & 1393.755 & 2&$> 14.75$&$> 14.92$&2&$> 14.61$&$> 14.60$\\
\ion{Si}{4} & 1402.770 & 2&$> 14.92$&&2&$> 14.60$&\\
\ion{P}{2} & 1152.818 &  & & $< 14.33$&3&$> 14.14$&$> 14.14$\\
\ion{P}{2} & 1532.533 & 5&$< 14.33$&&5&$< 14.14$&\\
\ion{S}{2} & 1250.584 &  & & $ 15.29 \pm 0.04$&1&$ 15.36 \pm 0.02$&$ 15.35 \pm 0.01$\\
\ion{S}{2} & 1253.811 & 1&$ 15.29 \pm 0.04$&&1&$ 15.35 \pm 0.01$&\\
\ion{Cr}{2} & 2056.254 & 5&$< 12.97$&$< 12.97$&1&$ 13.18 \pm 0.08$&$ 13.18 \pm 0.08$\\
\ion{Cr}{2} & 2066.161 &  & & &5&$< 13.12$&\\
\ion{Fe}{2} & 1608.451 & 3&$> 14.86$&$ 15.28 \pm 0.13$&3&$> 15.09$&$ 14.99 \pm 0.10$\\
\ion{Fe}{2} & 1611.200 & 1&$ 15.28 \pm 0.13$&&9&$ 14.99 \pm 0.10$&\\
\ion{Ni}{2} & 1370.131 &  & & $ 13.75 \pm 0.09$&1&$ 13.88 \pm 0.03$&$ 14.05 \pm 0.04$\\
\ion{Ni}{2} & 1454.842 & 1&$ 13.66 \pm 0.08$&&1&$ 13.89 \pm 0.04$&\\
\ion{Ni}{2} & 1709.604 &  & & &1&$ 14.08 \pm 0.04$&\\
\ion{Ni}{2} & 1741.553 & 1&$ 13.75 \pm 0.09$&&1&$ 14.11 \pm 0.03$&\\
\ion{Ni}{2} & 1751.916 &  & & &1&$ 14.05 \pm 0.04$&\\
\ion{Zn}{2} & 2026.136 & 5&$< 12.62$&$< 12.62$&1&$ 12.89 \pm 0.04$&$ 12.89 \pm 0.04$\\
\tableline
\end{tabular}
\end{center}
\tablenotetext{a}{Flag:  0=Non-dominant ion or blended transition.
1=Dominant ion; 2=Saturated, non-dominant ion; 3=Saturated, dominant ion
4=Upper limit, non-dominant ion; 5=Upper limit, dominant ion.}
\tablenotetext{b}{Weighted mean or most stringent limit.}
\tablecomments{Column densities are measured from the apparent 
optical depth method integrated over the entire line-profile.}
\end{table*}

The metallicity of the SLLS may be derived by comparing the metal
column densities against \nhi.  In gas that is predominantly neutral
(e.g.\ the majority of DLA), an accurate metallicity
value is calculated 
through a direct comparison of low-ion species, 
e.g.\ Si/H $\approx {\rm Si^+/H^0}$.  In systems with predominantly
photoionized gas, however, one must consider ionization corrections
to the low-ion ratios \citep{howk_ion99,vladilo01,pro_ion02}.
To gauge the ionization state, ideally one compares multiple
ionization levels of a single element.  Unfortunately, the
relatively low redshift of these SLLS places several
key transitions (e.g.\ \ion{Fe}{3}~1122, \ion{N}{2}~1083) below
the atmospheric cutoff.  One can, however, make a first crude 
estimate based on the \nhi\ values.  At $z \sim 2$, the extragalactic
background radiation field is sufficiently intense to photoionize the
majority of absorbers with $\mnhi < 10^{19.5}$
\citep{viegas95,prochaska96}.  Therefore, we expect that \slla\
is predominantly ionized with ionization fraction 
$x \equiv {\rm H^+/H} > 0.5$, perhaps with an ionization state
comparable to the SLLS at $z=2.65$ toward Q2231--00 
\citep[$x=0.99$;][]{prochaska99}.
At $\mnhi = 10^{20.10}$, \sllb\ is nearly a DLA and one expects
it to have a much lower ionization fraction than \slla.  

The observations place an additional constraint on the ionization
fraction.
First, we have measured $\N{Al^{++}}$ values from the
\ion{Al}{3}~1862 transitions and set an upper limit on the 
Al$^{++}$/Al$^+$ ratios based on the saturated \ion{Al}{2}~1670 profile
(Table~2).  These values, unfortunately, are not very
constraining and allow for both large and small $x$ values.
\cite{vladilo01} have shown that the Al$^{++}$/Si$^+$ ratio is
a useful diagnostic of photoionization.  The observed Al$^{++}$/Si$^+$
value for \slla\ is twice as large as that for any DLA
whereas the value for \sllb\ is comparable to that for DLA with
low \nhi\ value
\citep{vladilo01}.  Therefore, the values are consistent with a 
significant ionization fraction for \slla\ and predominantly neutral
gas for \sllb.  As such, we estimate the ionization corrections for
\sllb\ are negligible.

For \slla, we have considered the ionization corrections in greater
depth.  Specifically, we have performed a series of Cloudy 
calculations \citep[v05.07.06][]{ferland03} assuming a constant density
cloud with $\mnhi = 10^{18} \cm{-2}$ to $10^{19} \cm{-2}$, solar metallicity
and two input radiation fields: (i) an updated extragalactic UV background (EUVB)
model \citep[HM;][]{haa96} and (ii) the output from a Starburst~99 
calculation \citep[SB99;][]{sb99} for a galactic starburst with 
age of 100\,Myr.
As discussed below, the latter radiation field is motivated by the
fact that these SLLS absorbers resemble the gas associated with  
bright Lyman break galaxies \citep{steidel99,prs02}.
We considered a range of intensities parameterized by the ionization
parmeter $U \equiv \phi/n_H$ with $\phi$ the number density of ionizing
photons.  For the HM model and $\mnhi = 10^{19} \cm{-2}$, the
upper limit to Al$^{++}$/Al$^+$ implies upper limits to the ionization
corrections for Si$^+$ and Fe$^+$ of 0.6\,dex and 0.2\,dex respectively.
We note, however, that the \ion{Al}{2}~1670 profile is highly saturated
and a more reasonable upper limit to $\log({\rm Al^{++}/Al^+})$ is 
$-0.3$\,dex implying ionization corrections of less than 0.3\,dex for
all of the low-ions considered here.  To be conservative, we have adopted
0.3\,dex corrections, i.e.\ (X/H) = (X$^i$/\ion{H}{1})~$- 0.3$, 
for all low-ions X$^i$.  

If one were to assume a lower \nhi\ value for \slla, the 
ionization correction is larger but the corrected gas metallicity
is also larger (the increased ionization correction is 
smaller than the decrease in \nhi).
Similarly, if we adopt the S99 spectrum which is softer than the EUVB field
one also finds slightly larger corrections (0.1 to 0.2\,dex).  However,
the corrections for $\log({\rm Al^{++}/Al^+}) \approx -0.3$ are still
only $\approx 0.3$\,dex for Si$^+$ and S$^+$ relative to \ion{H}{1}.
We caution again that one should avoid drawing firm conclusions on the
ionization state of the gas from Al$^{++}$ and Al$^+$ alone because
of uncertainties in the recombination rates of these ions.
Unfortunately, a more accurate assessment of the ionization corrections must 
await observations at $\lambda < 3000$\AA.

\begin{table}[ht]\footnotesize
\begin{center}
\caption{{\sc SUMMARY\label{tab:summ}}}
\begin{tabular}{ccc}
\tableline
\tableline
&J0927+5621 & J0953+5230 \\
\tableline
z$_{\rm SLLS}$&1.775&1.768\\
\nhi & 19.00$^{+0.10}_{-0.25}$& 20.10$^{+0.10}_{-0.10}$ \\
$\lbrack$Si/H]$^a$ & $+ 0.72$&$+ 0.01$\\
$\lbrack$S/H] & $+ 0.79$&$+ 0.05$\\
$\lbrack$Zn/H] & $<+ 0.65$&$+ 0.12$\\
$\lbrack$S/Fe] & $+ 0.31$&$+ 0.66$\\
log(Al$^{++}$/Al$^+$)& $< 0.02$&$<-0.27$\\
log(Al$^{++}$/Si$^+$)& $-1.53$&$-1.81$\\
$x^b$ & 0.9 & $<0.1$ \\
\tableline
\end{tabular}
\end{center}
\tablenotetext{a}{[X/Y] is the logarithmic gas-phase abundance of species 
X relative to Y \\
relative to Solar abundance.  We have assumed low-ion 
species and \\
ionization corrections of 0.3\,dex for \\
\slla\ and 0\,dex for \sllb.}
\tablenotetext{b}{Adopted value based on inferences made from the
 observed \\
\nhi\ and Al$^{++}$ values.}
\end{table}

Table~2 presents the gas-phase abundances relative to solar
\citep{solarabnd}
based on the low-ion ratios with our adopted ionization corrections.
The Zn\footnote{Note that Zn is a trace element and even a $100\times$
enhancement would not require a large metallicity.},
Si, and S values for \slla\
are all consistent with the absorber having $\approx 5\times$ solar
abundance.  At present, the gas has the highest, precise metallicity 
measurement in any astrophysical environment 
\citep[e.g.][]{dhs03,jbt05,peroux06,gbc06,ptv06}.   
Again, we note that the
\ion{H}{1} column density could be significantly less than the value
adopted here. If we adopted a lower \nhi\ value we would derive a higher
ionization fraction and ionization correction, yet also a higher
gas metallicity. 
For \sllb, the metallicity is marginally super-solar.  
The observations demonstrate that super-solar gas exists at high
redshift, even apart from the direct vicinity of quasars.


\section{Discussion}

Before commenting on the implications for metals in the young
universe, let us consider the physical origin of this gas.
It is notable that both absorbers exhibit relatively wide absorption
line profiles revealing a velocity field of several hundred \kms.
While this is partly the effect of selection bias (large velocity
width yields larger EW), the MSDLA candidate sample \citep{herbert06}
focused primarily on \ion{Si}{2}~1808 and \ion{Zn}{2}~2026 which are 
optically thin in these systems. 
We suggest that the kinematics are indicative of 
feedback processes correlated with star formation.  Indeed, 
these absorbers may represent sightlines which pass through
the gas observed in absorption in the spectra of bright Lyman
break galaxies \citep{steidel99,prs02}.
The systems also exhibit very large \ion{C}{4} column density.
Current measurements of the frequency distribution of \ion{C}{4}
column densities show a dependence $N_{\rm CIV}^{-\alpha}$
with $\alpha < 2$ to \ion{C}{4} column densities of
$\mnciv = 10^{15} \cm{-2}$ \citep{songaila05}.  
Therefore, the mass density of C$^{+3}$ ions is dominated by
the largest column density absorbers.  We suspect that LLS like the
ones presented here
contain the majority of C$^{+3}$ ions at all redshifts.
Again, this is an assertion we will test through a large sample of 
LLS observations.
It would be valuable to compare the observations of the super-solar SLLS
with models of outflows from star-forming galaxies.

The detection of two super-solar SLLS underscores the prospect for
LLS to contain a substantial fraction of the metal mass
in the high $z$ universe.  Consider the results for \slla\ alone.
The total metal column density of the gas is 
given by $\log N_M = \log N_H + \log(M/H)$
where $N_H$ is the total hydrogen column density and M/H is the mean
metallicity of the gas.
Taking $x = 0.9$ ($N_H = 10 \mnhi$) and  
five times solar abundance, we find that the
metal column density is $50\%$ larger than any DLA at $z \approx 2$.
In fact, the $N_M$ value roughly matches the total of 
the $\approx 20$ DLA with metallicity measurements
at $z=1.6$ to 2.2 \citep{pro_mtl03}.
Because SLLS outnumber DLA by $\approx 4$ times \citep{omeara06}, 
even if systems like \slla\
represent only 1\% of the SLLS population, their contribution would
match the entire DLA population.   This remains true 
even if the other 99\% of SLLS were primordial!

To better illustrate the contribution of LLS (specifically SLLS)
on the cosmological metal budget, consider the following calculation.
First, define the mass density of metals in the DLA relative to the
critical density

\begin{equation}
\Omega_m^{DLA} = \Omega_{HI}^{DLA} \bar Z_{DLA}
\end{equation}
where $\bar Z_{DLA}$ is the mean metallicity in mass units
(i.e.\ $Z_\odot = 0.022$ is the Solar metallicity).
Adopting the results from 
\cite{phw05} $\Omega_{HI}^{DLA} = 0.5\sci{-3}$ and \cite{pro_mtl03} 
$\bar Z_{DLA} = 0.0022$,
we derive $\Omega_m^{DLA} = 1.1 \sci{-6}$, i.e.\ a few percent of \omt.

Consider the metal mass density of the SLLS, $\Omega_m^{SLLS}$, 
motivated by observational considerations:
\begin{equation}
\Omega_m^{SLLS} = \int \frac{1}{1-x} \mnhi f(\mnhi) \bar Z_{SLLS} \; d\mnhi
\label{eqn:mlls}
\end{equation}
where $x$ is the ionization fraction of the LLS (likely a function
of \nhi), $f$ is the frequency distribution, and 
$\bar Z_{LLS}$ is the mean metallicity of the gas.  
Currently, all of the expressions in the integrand of Equation~\ref{eqn:mlls}
are poorly constrained and so we will proceed conservatively.
First, we adopt our new measurement of $\mfnhi$ for the SLLS 
at $z \approx 2.5$ \citep{omeara06}: $\mfnhi = 10^{7.17} \mnhi^{-1.4}$. 
Second, we will adopt a value to the mean metallicity based
on the \cite{peroux_slls03} 
survey ($\bar Z_{SLLS} = 0.1 Z_\odot$) and assume the
value is independent of \nhi\ value.
In light of the results presented in this paper, we
consider this to be a conservative lower limit to $\bar Z_{SLLS}$.  
Even though super-solar SLLS are very
rare they will contribute significantly to the mean if they
represent $>1\%$ of the sample.

The most uncertain quantity in this calculation is the
ionization fraction which undoubtedly varies with \nhi\ column density
(and possibly metallicity).  Photoionization calculations estimate that
$x \approx 0$ at the DLA threshold and increases to $\approx 0.9$ at
$\mnhi = 10^{19} \cm{-2}$ with strongest dependence on the 
gas density and intensity
of the radiation field \citep[e.g.][]{vladilo01,pro_ion02}.
\cite{peroux_slls03} and \cite{mirka03} stressed that ionization
corrections are small for their SLLS sample and gave the misleading
impression that the gas is predominantly neutral.  Indeed, this
is not the case (Dessauges-Zavadsky, priv.\ comm.); we also
find that the SLLS can be significantly ionized 
(Prochaska 1999; Prochter et al.\, in prep.).
For the following, we will 
parameterize $x$ with an empirical motivation:  
$x = C_x (1 - \mnhi/10^{20.3})$
where $C_x$ can be varied to reduce/increase the degree of photoionization.

Evaluating Equation~\ref{eqn:mlls}, we derive 
$\Omega_m^{SLLS} = 5\sci{-6} = 0.15 \momt$
for $C_x=1$ and $2\sci{-6}$ for $C_x = 0.5$.  These values are 
significantly in excess of the metal mass density implied by the damped \lya\
systems and the IGM. 
We believe this calculation is reasonably conservative because
(i) we have neglected the contribution of LLS with $\mnhi < 10^{19} \cm{-2}$
and
(ii) the mean metallicity of the SLLS could easily be 1/3 or 1/2 solar.
Therefore, the LLS may contain the majority of metals at high $z$.
In future papers, we will derive empirical measurements of the
ionization fraction and metallicity of a large sample of LLS
to directly determine the LLS contribution.

\acknowledgments

The authors wish to recognize and acknowledge the very significant
cultural role and reverence that the summit of Mauna Kea has always
had within the indigenous Hawaiian community.  We are most fortunate
to have the opportunity to conduct observations from this mountain.
The authors wish to thank P. Madau, A. Aguirre, and R. Simcoe for valuable 
comments.
JXP and GEP are supported by NSF grant AST-0307408.

\end{document}